\documentclass[twocolumn,floatfix,superscriptaddress,a4paper,showpacs,showkeys,nofootinbib,reprint]{revtex4-1}
\usepackage{epsfig}
\usepackage{latexsym}
\usepackage{xspace}
\usepackage[colorlinks=true,linktocpage=true,linkcolor=blue,citecolor=blue,allcolors=blue]{hyperref}
\usepackage{url}
\usepackage[utf8]{inputenc}
\usepackage{indentfirst}
\usepackage{enumerate}
\usepackage{color}

\usepackage[caption=false,position=top]{subfig}

\usepackage{amsmath}
\usepackage{amssymb}
\usepackage[english]{babel}
\usepackage{url}

\begin{document}


\title{
Kinetic freeze-out temperature from 
yields of short-lived resonances 
}

\author{Anton Motornenko}
\affiliation{
Institut f\"ur Theoretische Physik,
Goethe Universit\"at Frankfurt, Max-von-Laue-Str. 1, D-60438 Frankfurt am Main, Germany}
\affiliation{Frankfurt Institute for Advanced Studies, Giersch Science Center, Ruth-Moufang-Str. 1, D-60438 Frankfurt am Main, Germany}

\author{Volodymyr Vovchenko}
\affiliation{
Institut f\"ur Theoretische Physik,
Goethe Universit\"at Frankfurt, Max-von-Laue-Str. 1, D-60438 Frankfurt am Main, Germany}
\affiliation{Frankfurt Institute for Advanced Studies, Giersch Science Center, Ruth-Moufang-Str. 1, D-60438 Frankfurt am Main, Germany}
\affiliation{
Nuclear Science Division, Lawrence Berkeley National Laboratory, 1 Cyclotron Road, Berkeley, CA 94720, USA}

\author{Carsten Greiner}
\affiliation{
Institut f\"ur Theoretische Physik,
Goethe Universit\"at Frankfurt, Max-von-Laue-Str. 1, D-60438 Frankfurt am Main, Germany}

\author{Horst Stoecker}
\affiliation{
Institut f\"ur Theoretische Physik,
Goethe Universit\"at Frankfurt, Max-von-Laue-Str. 1, D-60438 Frankfurt am Main, Germany}
\affiliation{Frankfurt Institute for Advanced Studies, Giersch Science Center, Ruth-Moufang-Str. 1, D-60438 Frankfurt am Main, Germany}
\affiliation{GSI Helmholtzzentrum f\"ur Schwerionenforschung GmbH, D-64291 Darmstadt, Germany}

\begin{abstract}
A method to determine the kinetic freeze-out temperature in heavy-ion collisions from measured yields of short-lived resonances is presented.
The resonance production
is treated
in the framework of thermal model with an evolution between chemical and kinetic freeze-outs.
The yields of many short-lived resonances are suppressed at $T = T_{\rm kin} < T_{\rm ch}$.
We determine the values of $T_{\rm kin}$ and $T_{\rm ch}$ for various centralities in Pb--Pb collisions at $\sqrt{s_{_{NN}}} = 2.76$~TeV by fitting the abundances of both the stable hadrons and the short-lived resonances such as $\rho^0$ and $ \text{K}^{*0}$, that were measured by the ALICE collaboration.
This allows to extract the kinetic freeze-out temperature from the measured hadron and resonance yields alone,
independent of assumptions about the flow velocity profile and the freeze-out hypersurface.
The extracted $T_{\rm ch}$ values exhibit a moderate multiplicity dependence whereas $T_{\rm kin}$ drops, from $T_{\rm kin} \simeq T_{\rm ch} \simeq 155$~MeV in peripheral collisions to $T_{\rm kin} \simeq 110$~MeV in 0-20\% central collisions. 
Predictions for other short-lived resonances are presented.
A potential (non-)observation of a suppressed $f_0(980)$ meson yield will allow to constrain the lifetime of that meson.

\end{abstract}



\maketitle


\section{Introduction}

Relativistic heavy-ion experiments at the Schwerionen Synchrotron~(SIS), the Super Proton Synchrotron~(SPS), the Relativistic Heavy Ion Collider~(RHIC), and the Large Hadron Collider~(LHC) provide a rich dataset of spectra and abundances of identified particles~\cite{Abelev:2013vea,Adamczyk:2017iwn}. This includes both, the long-lived and short-lived hadrons.
The abundances of stable hadrons agree quite well with a thermal model calculation, characterized by the chemical freeze-out temperature $T_{\rm ch} \simeq 150-160$~MeV~\cite{Becattini:2012xb,Petran:2013lja,Andronic:2017pug,Adamczyk:2017iwn}.
The yields of short-lived resonances, like $\text{K}^*$ or $\rho$, are significantly overpredicted by the thermal model~\cite{Aggarwal:2010mt,Anticic:2011zr,Abelev:2014uua,Acharya:2018qnp}, indicating a presence of an additional mechanism which suppresses these yields.
This suppression is attributed to the existence of a hadronic phase. The expanding system seems to maintain kinetic (but not chemical) equilibrium after the chemical freeze-out down to a kinetic freeze-out temperature $T_{\rm kin} < T_{\rm ch}$.

The kinetic freeze-out temperature has often been extracted from blast-wave fits to the $p_T$ spectra of stable hadrons.
This procedure assumes an interplay of a particular flow velocity profile and a kinetic freeze-out hypersurface.
Cylindrically-symmetric blast-wave models are often used~\cite{Schnedermann:1993ws},
which yield $T_{\rm kin} \sim 100$~MeV for the most central collisions at LHC~\cite{Adam:2015vda}, RHIC~\cite{Adamczyk:2017iwn}, and SPS~\cite{Anticic:2016ckv}.
However, different freeze-out geometries can lead to different conclusions~\cite{Broniowski:2001we}.
Here we present a novel procedure on how to extract $T_{\rm kin}$. 
The method is independent of assumptions about the flow velocity profile and the freeze-out hypersurface. 

\section{Methodology}

The observed suppression of ``thermal'' resonance yields is usually attributed to rescattering of the decay products in the hadronic phase~\cite{Knospe:2015nva,Cho:2015exb,Steinheimer:2017vju}. Then these short-lived resonances can no longer be identified in invariant mass measurements. Hence, this looks like the ``observed'' resonance yields are suppressed. 
Such a picture has been used previously to estimate the lifetime of the hadronic phase at RHIC and SPS energies from the measured resonance abundances~\cite{Rafelski:2001hp,Torrieri:2001ue,Markert:2002rw}, neglecting the effect of resonance regeneration.
The scattering cross sections of various elastic meson-meson and meson-baryon reactions, however, are in fact dominated by the formation of intermediate short-lived resonance states~\cite{Tanabashi:2018oca}.
Common examples are $\pi \pi \to \rho \to \pi \pi$, $\pi \text{K} \to \text{K}^* \to \pi \text{K}$, and $\pi \text{N} \to \Delta \to \pi \text{N}$.
Rescattering of a resonance decay product is likely to regenerate a resonance.
Transport model calculations~\cite{Steinheimer:2017vju} indeed show that repeated resonance-formation dominates pure elastic meson-meson and meson-baryon rescatterings in the hadronic phase.

The resonance-forming pseudo-elastic reactions obey the law of mass action during the hadronic phase. They are the primary driver for maintaining the kinetic equilibrium in expanding systems, and lead to the following scenario:
\begin{itemize}
    \item At the chemical freeze-out, at $T = T_{\rm ch}$, the inelastic reaction rates drop out of equilibrium. 
    The total yields of all stable hadrons become frozen.
    The total hadron yield corresponds to the sum of the yields of primordial hadrons and those which stem from decays of short-lived resonances.
    The final abundances of stable hadrons are described by the standard chemical equilibrium thermal model.
    
    \item The system then expands and cools isentropically, until the kinetic freeze-out temperature $T_{\rm kin} < T_{\rm ch}$ is reached.
    This stage is identified with the hadronic phase. It is modeled by a concept of partial chemical equilibrium~(PCE)~\cite{Bebie:1991ij}.
    The decays and the regenerations of the short-lived resonances obey the law of mass action, i.e. the abundances of the different resonances stay in equilibrium with those particles which are formed in the decays of these resonances.
    
    \item The remaining resonances then decay after the kinetic freeze-out. Their decay products do not rescatter and the resonance regeneration ceases to occur. The resonance abundances at $T = T_{\rm kin}$ are identified with those measured experimentally.
    This implies that the chemical freeze-out of short-lived resonances coincides with the kinetic freeze-out of bulk hadron matter.

\end{itemize}
Of course, the actual decoupling of particles in an expanding system is a continuous process that takes place over a range of temperatures. 
In that sense the $T_{\rm ch}$ and $T_{\rm kin}$ temperatures characterize average conditions for the chemical and kinetic freeze-outs.

The present scenario is largely consistent with the seminal ideas regarding strangeness production in heavy-ion collisions~\cite{Koch:1986ud}.
Quantitatively, the thermodynamic properties of the system in the hadronic phase are described here using a hadron resonance gas~(HRG) model in PCE~\cite{Bebie:1991ij,Hung:1997du,Hirano:2002ds,Kolb:2002ve,Huovinen:2007xh}.
The effective chemical potentials $\tilde{\mu}_j$ of all species are thus given by
\begin{equation}
\label{eq:mus}
\tilde{\mu}_j = \sum_{i \in \rm stable} \, \langle n_i \rangle_j \, \mu_i~.
\end{equation}
The index $i$ runs over all particles, whose final abundances are frozen at $T = T_{\rm ch}$. These hadrons are identified with the ones stable w.r.t. strong decays, i.e. $\pi$, N, $\eta$, $\eta'$, K, $\Lambda$, $\Sigma$'s, $\Xi$'s, $\Omega$, as well their antiparticles\footnote{Alternatively, one can treat the yields of long-lived resonances such $\phi$, $\omega$, $\Xi(1530)$, and/or $\Lambda(1520)$ to be frozen at $T = T_{\rm ch}$ as well~\cite{Huovinen:2007xh}. We verified that the results presented here look very similar in such a scenario.}.
$\mu_i$ are the chemical potentials of particles considered stable. $\langle n_i \rangle_j$ is the mean number of hadron species $i$ resulting from decays of hadron species $j$.
The PCE evolution of the system follows from the conditions of the conservation of the total yields of the stable hadrons as well as of the entropy:
\begin{align}
\label{eq:pce:Ni}
\sum_{j \in {\rm HRG}} \langle n_i \rangle_j \, n_j(T, \tilde{\mu}_j) \, V & = N_i^{\rm tot}(T_{\rm ch}), ~~ i \in \rm stable,  \\
\label{eq:pce:S}
\sum_{j \in {\rm HRG}} s_j(T, \tilde{\mu}_j) \, V & = S (T_{\rm ch})~.
\end{align}
These equations provide the chemical potentials $\mu_j$ and the system volume $V$ during the system's expansion.
The index $j$ runs over all hadrons and resonances in the list, $n_j$ and $s_j$ are the grand-canonical number- and entropy densities of the hadron species $j$ in the multi-component ideal hadron gas, $N_i^{\rm tot}(T_{\rm ch})$ and $S(T_{\rm ch})$ are, respectively, the total yield of stable hadron species $i$ and the total entropy of the system during the whole expansion.
We use the energy-dependent Breit-Wigner~(eBW) scheme for modeling the spectral functions of all resonances~\cite{Vovchenko:2018fmh}.
On the other hand, the energy dependence of branching ratios is neglected. The PDG branching ratios are used throughout to evaluate $\langle n_i \rangle_j$.
Excluded-volume and strangeness undersaturation effects are omitted unless stated otherwise.
The calculations are performed using the open source \texttt{Thermal-FIST} package~\cite{Vovchenko:2019pjl}, which contains a numerical implementation of the PCE-HRG model defined above~(available since version 1.3 via~\cite{FIST}).

The numerical solution of Eqs.~\eqref{eq:pce:Ni} and \eqref{eq:pce:S} yields the temperature dependence of the volume $V$ and of the chemical potentials $\tilde{\mu}_j$ of all the species during the hadronic phase.
Within our PCE-HRG model implementation, this dependence was presented in Ref.~\cite{Vovchenko:2019aoz} for the LHC energies.
The yield ratios involving short-lived resonances, such as $\text{K}^{*} / \text{K}$ and $\rho/\pi$, are not conserved during the hadronic phase. 
They decrease as the system cools down, 
their values at $T = T_{\rm kin}$ possibly describing the suppression seen in measurements, as first predicted in Ref.~\cite{Rapp:2003ar} long before precision data were available.
This is used here to extract the kinetic freeze-out temperature from experimental data.

\section{Data analysis}

The kinetic freeze-out temperature is determined for 2.76~TeV Pb--Pb collisions at the LHC by performing PCE-HRG model fits to the measured yields of pions, kaons, protons, $\Lambda$, $\Xi$, $\Omega$, $\phi$, $\text{K}_0^S$, $\text{K}^{*0}$, and $\rho^0$, of the ALICE collaboration, for 0-20\%, 20-40\%, 40-60\%, and 60-80\% centralities~\cite{Abelev:2013vea,Abelev:2013xaa,ABELEV:2013zaa,Abelev:2014uua,Acharya:2018qnp}. The yields are symmetrized between particles and antiparticles, i.e. we assume $\mu_B = 0$.
Three parameters of the fit are employed: the chemical freeze-out temperature $T_{\rm ch}$ and volume $V_{\rm ch}$, and the kinetic freeze-out temperature $T_{\rm kin}$.
The final yields of all species are evaluated at $T = T_{\rm kin}$.
The single freeze-out scenario, $T_{\rm kin} = T_{\rm ch}$, is also analyzed. All the abundances of all species are, in this single freeze-out scenario, described by the chemical equilibrium ideal HRG model.
The PCE-HRG fit procedure described above has been implemented in \texttt{Thermal-FIST} since version 1.3 and can be obtained via Ref.~\cite{FIST}.

\begin{figure}[t]
  \centering
  \includegraphics[width=.48\textwidth]{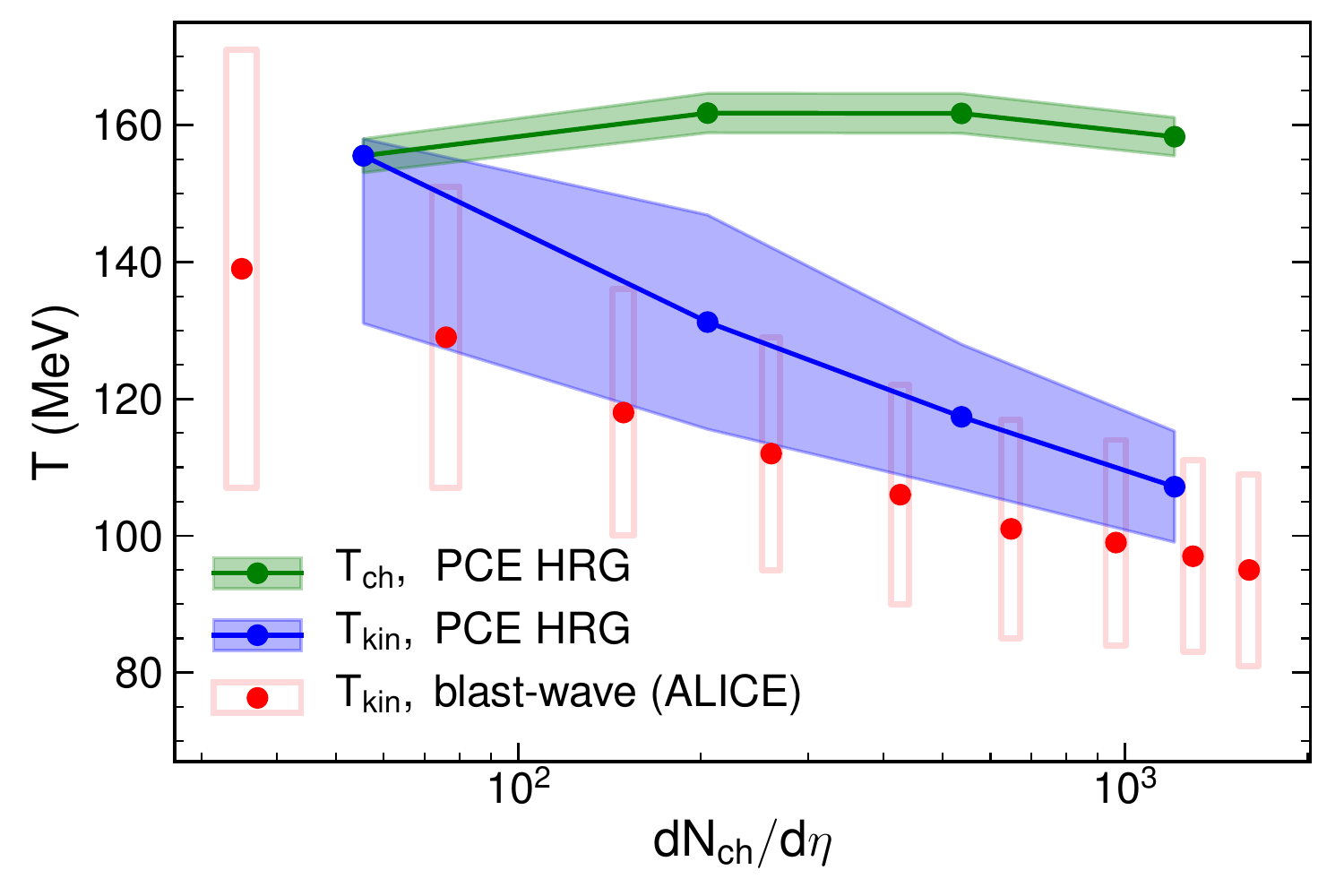}
  \caption{
  The values of the chemical $T_{\rm ch}$~(green symbols) and kinetic $T_{\rm kin}$~(blue symbols) freeze-out temperatures extracted from the PCE-HRG model fits to the ALICE collaboration data on the production of hadrons and resonances in Pb+Pb collisions at $\sqrt{s_{\rm NN}}=2.76$ TeV for various centralities, 
  depicted as a function of charged multiplicity.
  The red symbols depict the $T_{\rm kin}$ values extracted from blast-wave fits to the $p_T$ spectra of pions, kaons, and protons in Ref.~\cite{Abelev:2013vea}.
  }
  \label{fig:Tch-Tkin}
\end{figure}

The fit results are exhibited in Table~\ref{tab:fits}.
The centrality dependencies of both $T_{\rm kin}$ and $T_{\rm ch}$ are shown in Fig.~\ref{fig:Tch-Tkin} as a function of the charged particle multiplicity $dN_{\rm ch} / d\eta$~\cite{Aamodt:2010cz}. 
Figure 2 depicts the resulting data-over-model ratios for all hadron species used in fits.
The fitting parameter errors are obtained by analyzing the $\chi^2$ profiles. 
The error bar of $T_{\rm kin}$ is asymmetric for the $60-80$\% centrality because of the restriction $T_{\rm kin} \leq T_{\rm ch}$.

\begin{table}
 \caption{Results of the PCE-HRG model thermal fits to ALICE data for $\sqrt{s_{_{NN}}} = 2.76$~TeV Pb-Pb collisions at different centralities.
 For each centrality the first row corresponds to the single freeze-out scenario while the second row corresponds to separate chemical and kinetic freeze-outs scenario.
 } 
 \centering                                                 
 \begin{tabular}{@{\extracolsep{10pt}}cccc}   
 \hline
 \hline
 Centrality & $T_{\rm ch}$~(MeV) & $T_{\rm kin}$~(MeV) & $\chi^2 / {\rm dof}$  \\
 \hline
 0-20\% & $160.2 \pm 3.1$ & -- & 23.6/8 \\
        & $158.3 \pm 2.8$ & $107.1 \pm 8.2$ & 10.5/7 \\
\hline
 20-40\% & $162.9 \pm 3.1$ & -- & 19.5/8 \\
        & $161.7 \pm 2.9$ & $117.3 \pm 10.8$ & 12.8/7 \\
\hline
 40-60\% & $162.3 \pm 3.0$ & -- & 12.5/8 \\
        & $161.8 \pm 2.9$ & $131.2 \pm 15.9$ & 10.6/7 \\
\hline
 60-80\% & $155.5 \pm 2.5$ & -- & 19.1/8 \\
        & $155.5 \pm 2.5$ & $155.5^{+2.5}_{-24.5}$ & 19.1/7 \\
\hline
\hline
 \end{tabular}
\label{tab:fits}
\end{table}

The single freeze-out scenario cannot describe simultaneously the yields of stable hadrons and short-lived resonances in central collisions.
The $\text{K}^{*0}$ and $\rho^0$ yields are significantly overestimated by the model with $T_{\rm ch}=T_{\rm kin}\simeq 155$~MeV.
That situation improves in peripheral collisions, where the apparent suppression of the resonance yields appears to be milder.
The separation of kinetic and chemical freeze-outs leads to an improved description of the measured yields for all centralities, except for the most peripheral bin. 
$T_{\rm ch}$ exhibits little centrality dependence, its value is consistent with 155-160~MeV range throughout.
The extracted kinetic temperature increases monotonically from $T_{\rm kin} \simeq 110$~MeV for the 0-20\% centrality bin, to $T_{\rm kin} \simeq T_{\rm ch} = 155$~MeV for 60-80\% centrality.
This result indicates the existence of a hadronic phase in heavy-ion collisions, a rather long-lived one in central collisions and a short-lived one in peripheral collisions.

\begin{figure}[t]
  \centering
  \includegraphics[width=.48\textwidth]{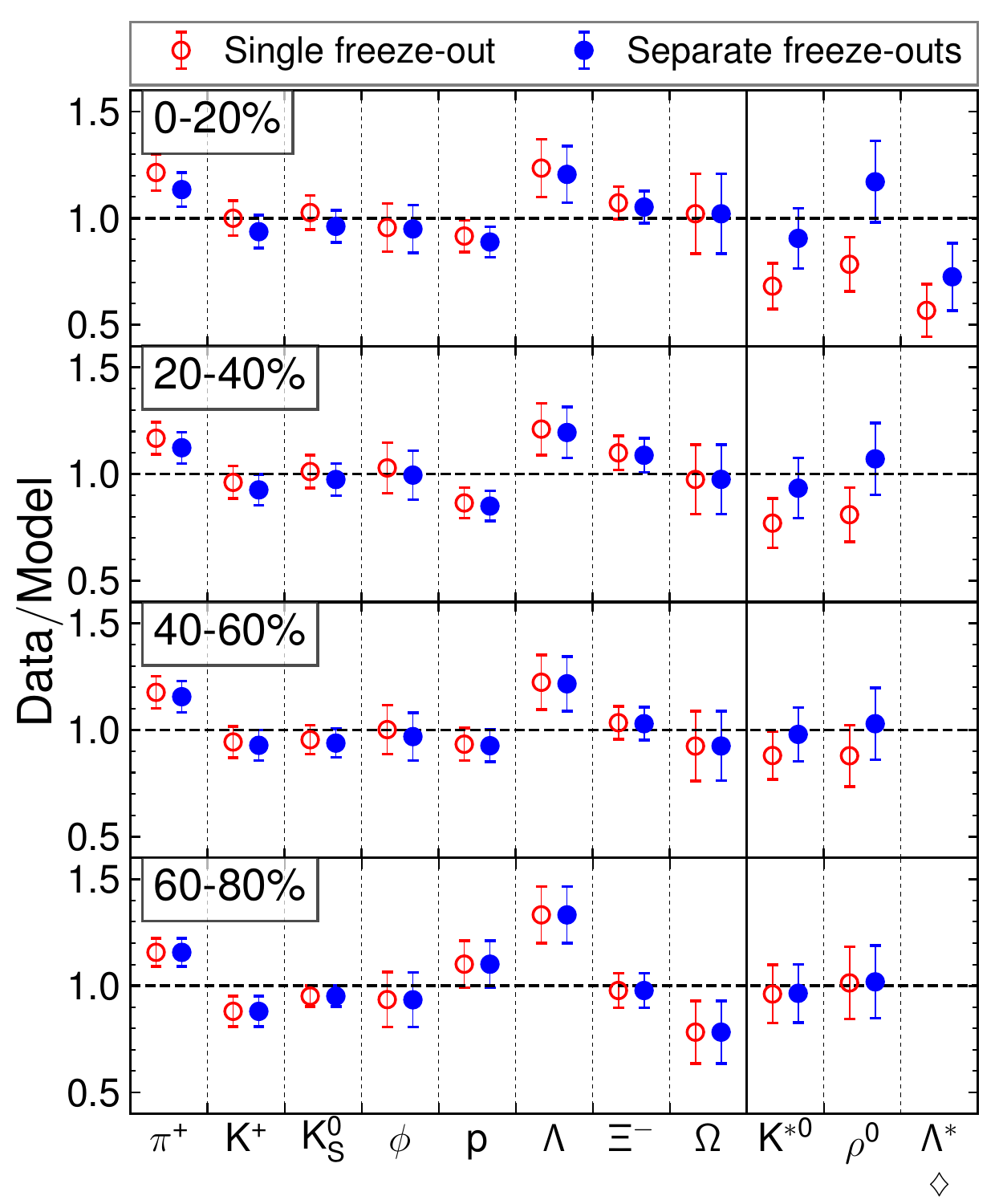}
  \caption{The data/model ratios resulting from thermal fits to particle yields measured in Pb-Pb collisions of various centrality at $\sqrt{s_{NN}} = 2.76$ TeV. Fits are performed within the single freeze-out HRG picture (open red circles), and the separate freeze-outs PCE-HRG picture (full blue circles).
  Here $\Lambda^*$ corresponds to the $\Lambda(1520)$. The $\Lambda(1520)$ yields were not used in the fit procedure.
  }
  \label{fig:model_data}
\end{figure}

Let us compare these results to the $T_{\rm kin}$ values resulting from blast-wave model fits~\cite{Schnedermann:1993ws} to the $p_T$ spectra of pions, kaons, and protons, as presented by the ALICE collaboration in Ref.~\cite{Abelev:2013vea}~(red symbols in Fig.~\ref{fig:Tch-Tkin}).
Our results are in fair agreement with this analysis, although the $T_{\rm kin}$ values of Ref.~\cite{Abelev:2013vea} are on the lower side of our error bands.
Recent blast-wave model studies~\cite{Mazeliauskas:2019ifr,Melo:2019mpn} take into account modifications of the $p_T$ spectra due to resonance feeddown.
The $T_{\rm kin}$ values of Ref.~\cite{Mazeliauskas:2019ifr} lie considerably closer to $T_{\rm ch}$ than in the present study, whereas Ref.~\cite{Melo:2019mpn} reports a much smaller value $T_{\rm kin} \simeq 80$~MeV for most central collisions.
This large spread of the $T_{\rm kin}$ values reported in the literature is an indication of significant systematic uncertainties which are currently present in the blast-wave model approach. 
It should be noted that none of those above three analyses does incorporate constraints from the data on short-lived resonances, in contrast to the study presented here. 
Thus, inclusion of the measured spectra of resonances is one way to improve the blast-wave approach.
The $p_T$ spectra fits also depend on the validity of the blast-wave model's assumed flow velocity profile and freeze-out hypersurface.
The concept presented here is free of this issue.

The systematic uncertainties associated with the implementation of the HRG model itself deserve attention. 
In addition to the eBW scheme, we considered also the zero-width treatment of resonances.
The extracted $T_{\rm ch}$ and $T_{\rm kin}$ values are, respectively, about 2-3 MeV smaller and 5 MeV larger in the zero-width case than in the eBW case.
The fit quality worsens for all centralities (except for the most peripheral bin). 
This is mainly a consequence of the increased proton yield in the zero-width scheme.
The effects of incomplete strangeness equilibration are studied by introducing a strangeness saturation parameter, $\gamma_S \leq 1$~\cite{Koch:1986ud,Rafelski:1991rh}.
This has a notable influence on the most peripheral~(60-80\%) bin only. Here, $\gamma_S \simeq 0.85$, and the extracted $\chi^2$ value decreases by about a factor of two, while $T_{\rm ch}$ increases to about 160~MeV.
This is in line with previous statistical model analyses of the LHC data~\cite{Becattini:2014hla,Sharma:2018jqf,Vovchenko:2019kes}.
Separate chemical freeze-outs of strange and non-strange hadrons is another possibility which has been discussed~\cite{Bellwied:2013cta}.

Excluded volume corrections are often incorporated into the HRG model, and can have a 
sizable influence on thermal fits to the data~\cite{Yen:1997rv,Vovchenko:2015cbk,Alba:2016hwx}.
A moderate excluded volume correction is considered here by repulsive (anti)baryon-(anti)baryon interactions with a baryonic eigenvolume parameter $b \simeq 1$~fm$^3$. This is motivated by the analysis of lattice QCD data on baryon number susceptibilities~\cite{Vovchenko:2017drx} and Fourier coefficients~\cite{Vovchenko:2017xad}.
The excluded-volume PCE-HRG model fits yield $T_{\rm ch}$ and $T_{\rm kin}$ values which are, respectively, about 2-3 MeV larger and 2-3 MeV smaller than in the ideal HRG case.
The changes in the $\chi^2$ values are insignificant.

Yields of light nuclei are often considered in the thermal model HRG approach~\cite{Andronic:2010qu}.
Within the PCE-HRG framework one can either treat the nuclei as stable species whose yields are frozen at $T_{\rm ch}$ or one accepts that these fragile objects can be destroyed and regenerated during the evolution in the hadronic phase. In the latter case the nuclear abundances are in equilibrium with the abundances of their constituents, as follows from the Saha equation~(see Ref.~\cite{Vovchenko:2019aoz} for details).
We verified that the available data on light nuclei production in 0-20\% central Pb--Pb collisions are well described in both scenarios.

\section{Predictions}

\begin{figure}[t]
  \centering
  \includegraphics[width=.48\textwidth]{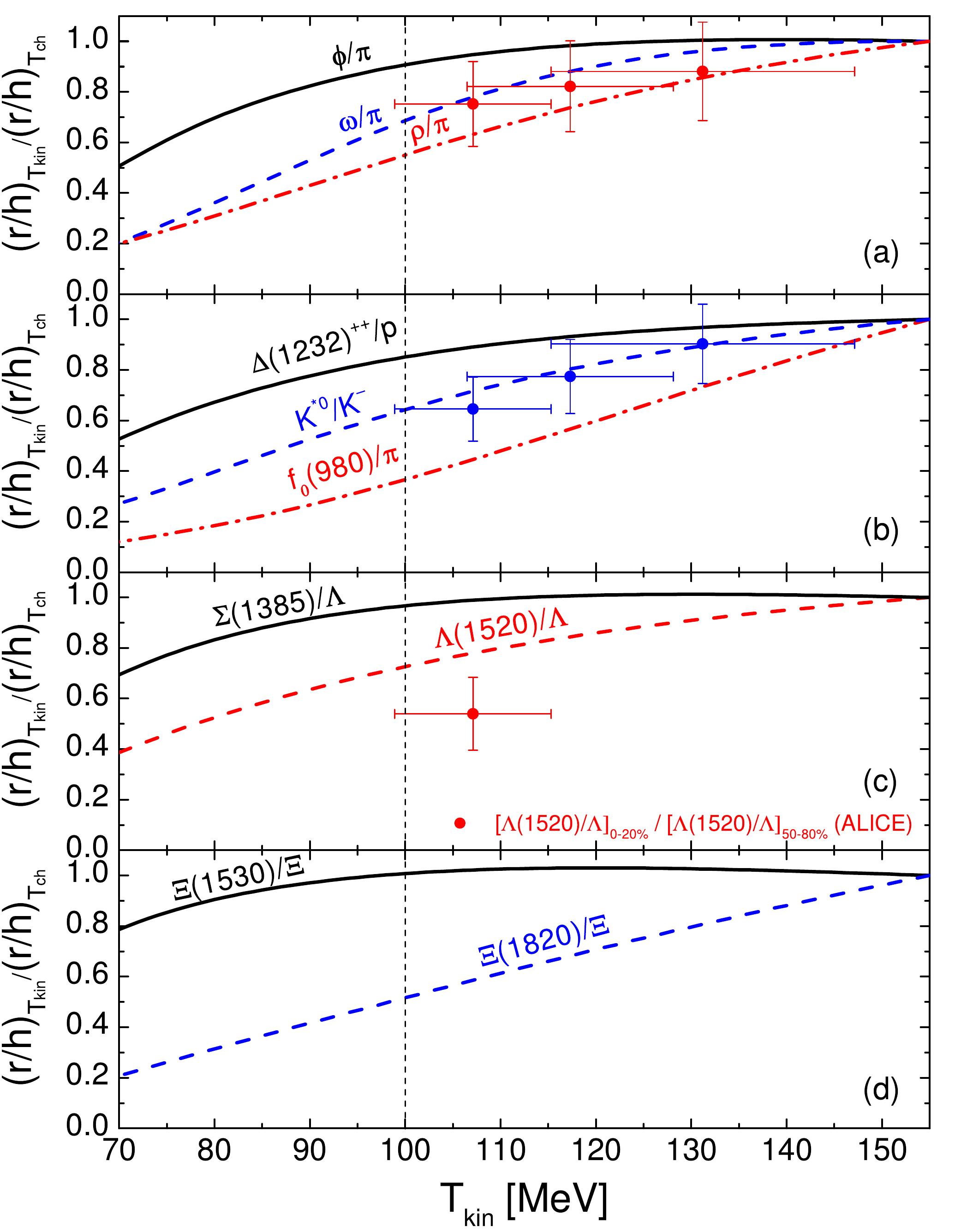}
  \caption{
  Dependence of the yield ratios (a) $2\phi / (\pi^+ + \pi^-)$~(solid black line), $2\omega / (\pi^+ + \pi^-)$~(dashed blue line), and $2\rho^0 / (\pi^+ + \pi^-)$~(dot-dashed red line), (b) $\Delta^{++}/\text{p}$~(solid black line), K$^{*0}$/K$^-$~(dashed blue line), and  $2f_0(980) / (\pi^+ + \pi^-)$~(dot-dashed red line), (c) $\Sigma(1385)/\Lambda$~(solid black line) and $\Lambda(1520)/\Lambda$~(dashed red line), and (d) $\Xi(1530)/\Xi$~(solid black line) and $\Xi(1820)/\Xi$~(dashed blue line) on the kinetic freeze-out temperature $T_{\rm kin}$.
  The ratios are normalized to their values at $T = T_{\rm ch} = 155$~MeV.
  The points in (a) and (b) depict, respectively, the experimental data for the ratios of $2\rho^0 / (\pi^+ + \pi^-)$ and K$^{*0}$/K$^-$ in 0-20\%, 20-40\%, and 40-60\% relative to the ones in 60-80\% Pb--Pb collisions at $\sqrt{s_{NN}} = 2.76$~TeV.
  The red point in (c) depicts ALICE collaboration data~\cite{ALICE:2018ewo} for the ratio of $\Lambda(1520)/\Lambda$ measured in 0-20\% to the one in
  50-80\% Pb--Pb collisions at $\sqrt{s_{NN}} = 2.76$~TeV.
  The dashed vertical line corresponds to $T_{\rm kin} = 100$~MeV, which is a typical value for the kinetic freeze-out temperature in most central collisions.
  }
  \label{fig:DoubleRatios}
\end{figure}

Various resonance-to-stable hadron yield ratios  can be analyzed in the PCE-HRG picture.
The resonances which can prospectively be measured are particularly interesting.
Specifically, the behavior of ratios $\phi / \pi$, $\omega / \pi$, $\rho / \pi$,   $\Delta^{++}/\text{p}$, $\text{K}^{*0} / \text{K}^-$, $f_0(980) / \pi$, $\Sigma(1385)/\Lambda$,
$\Lambda(1520)/\Lambda$, $\Xi(1530)^0/\Xi$, and $\Xi(1820)/\Xi$, is studied here at the LHC conditions~($\mu_B = 0$). 
Given the mild centrality dependence of the extracted chemical freeze-out temperature, here we fix $T_{\rm ch} = 155$~MeV and study the dependence of the ratios on $T_{\rm kin}$ only\footnote{The results are not sensitive to the specific value of $T_{\rm ch}$, e.g. $T_{\rm ch} = 160$~MeV gives very similar results.}.
All ratios are normalized by their values at $T_{\rm ch} = 155$~MeV in order to eliminate the influence of effects not related to the hadronic phase dynamics.
These double ratios quantify the suppression of resonance yields in (semi-)central collisions, where $T_{\rm kin} < T_{\rm ch}$, relative to the most peripheral collisions~(or, alternatively, to a pp/pA baseline), where $T_{\rm kin} \simeq T_{\rm ch}$. 
The $T_{\rm kin}$ dependence of the above-listed double-ratios is depicted in Fig.~\ref{fig:DoubleRatios}.

In general, different resonance particles can have different freeze-out temperatures, based on their interactions in the medium. Temperature dependence in Fig.~\ref{fig:DoubleRatios} allows to see the effect of the hadronic phase on different resonances depending on a chosen freeze-out temperature.
The yields of long-lived resonances $\phi$~($\tau \approx 46$~fm/$c$) and $\Xi(1530)$~($\tau \approx 22$~fm/$c$) change little in the PCE-HRG picture for $T_{\rm kin} \gtrsim 100$~MeV, even though this scenario assumes equilibrium of these long-lived resonances with their decay products.
Thus, an absence of suppression of the yields of these two resonances does not necessarily implies that these objects do not interact after the chemical freeze-out.
The long-lived $\omega$ meson~($\tau = 23$~fm/$c$), on the other hand, would be notably suppressed in central collisions if it interacts in the hadronic phase.

The short-lived $\rho^0$~($\tau = 1.3$~fm/$c$) and $\text{K}^{*0}$~($\tau = 4.2$~fm/$c$) meson yields exhibit a significant suppression as $T_{\rm kin}$ is lowered, as elaborated earlier.
On the other hand, the yields of short-lived baryonic resonances $\Delta^{++}$~($\tau = 1.7$~fm/$c$) and $\Sigma(1385)$~($\tau = 5$~fm/$c$) change only mildly.
For $T_{\rm kin} = 100$~MeV one observes only a 10-15\% suppression in the $\Delta^{++}$ yields and virtually no change for $\Sigma(1385)$.
Thus, if the mild system-size dependence of the yield ratios involving these resonances will indeed be observed experimentally, such an observation cannot be interpreted as an evidence against the existence of a long-lived hadronic phase.
The presented observations are qualitatively consistent with prior results of Monte Carlo simulations of heavy-ion collisions employing the hadronic afterburner UrQMD~\cite{Knospe:2015nva}.
The result for $\Sigma(1385)/\Lambda$ is also in line with a mild system-size dependence of this ratio observed at RHIC for $\sqrt{s_{NN}} = 200$~GeV~\cite{Adams:2006yu}.

A particularly interesting case is the scalar $f_0(980)$ meson.
The nature of $f_0(980)$ is not established and its lifetime is not constrained.
The PDG listing~\cite{Tanabashi:2018oca} gives $\Gamma_{f_0} \sim 10$--$100$~MeV.
This corresponds to a lifetime between about 2 and 20~fm/$c$.
In the former case, the lifetime is shorter than the lifetime of the hadronic phase and the PCE-HRG model assumption of detailed balance between decays and regenerations of $f_0(980)$ is justified.
Figure~\ref{fig:DoubleRatios} shows that the $f_0(980)/\pi$ ratio will be significantly suppressed in such a case if the hadronic phase is long-lived, e.g. the ratio drops by about a factor of three for $T_{\rm kin} = 100$~MeV.
On the other hand, if the $f_0(980)$ lifetime is large, then it is more reasonable to expect that its yield is frozen at $T_{\rm ch}$ and will not be modified appreciably in the hadronic phase.
The measurements of the $f_0(980)/\pi$ ratio in heavy-ion collisions at different centralities at the LHC~(or RHIC) can thus provide an indirect information on its lifetime:
A significant suppression of the $f_0(980)/\pi$ ratio in central collisions relative to peripheral ones can be interpreted as evidence for a short $f_0(980)$ lifetime.
An absence of such suppression, on the other hand, favors a large $f_0(980)$ lifetime. 
\\

\section{Summary and conclusions}

We developed a novel method to extract the kinetic freeze-out temperature in heavy-ion collisions based on the yields of short-lived resonances. 
This method, which employs hadron resonance gas model in partial chemical equilibrium, is agnostic to the assumptions regarding the flow velocity profile and the freeze-out hypersurface, that plague the commonly performed fits to the $p_T$ spectra.
The analysis of ALICE data on Pb--Pb collisions at the LHC yields a moderate multiplicity dependence of $T_{\rm ch}$ whereas the kinetic freeze-out temperature drops from $T_{\rm kin} \simeq T_{\rm ch} \simeq 155$~MeV in peripheral collisions to $T_{\rm kin} \simeq 110$~MeV in 0-20\% most central collisions. This result is in qualitative agreement with prior studies employing the blast-wave model fits.

Not all short-lived resonances exhibited a suppression of their yields due to a long-lasting hadronic phase: In contrast to $\rho^0$ and $\text{K}^{*0}$ mesons, the yields of baryon resonances  $\Delta^{++}$ and $\Sigma(1385)$ change little in the hadronic phase.
We point out 
a possibility to constrain the lifetime of $f_0(980)$ meson: 
A (non)observation of a suppressed $f_0(980)/\pi^{\pm}$ ratio in central collisions favors a long (short) $f_0(980)$ lifetime.
In the future we plan to extend our framework to lower collision energies, and also to analyze other sensitive probes of freeze-out dynamics, such as fluctuations and correlations of identified hadron numbers~\cite{Jeon:1999gr,Torrieri:2010te}.


\begin{acknowledgments}

We would like to thank F.~Bellini, P.~Braun-Munzinger, B.~D\"onigus, A.~Kalweit, A.~Mazeliauskas, and J.~Steinheimer for fruitful discussions.
We also thank S.~Cho and R.~Rapp for pointing out references pertinent to the present work.
H.St. acknowledges the support through the Judah M. Eisenberg Laureatus Chair at Goethe University by the Walter Greiner Gesellschaft, Frankfurt, and the BMBF programme ErUM, research field ``Universe and Matter''.

\end{acknowledgments}

\bibliography{TkinReso}


\end{document}